\documentstyle[prl,epsf, aps]{revtex}
\begin{document}
\newcommand{\ve}[1]{\mbox{\boldmath $#1$}} 

\draft
\twocolumn[\hsize\textwidth\columnwidth\hsize\csname @twocolumnfalse\endcsname

\title {Optical Confinement of a Bose-Einstein Condensate}

\author{D.M. Stamper-Kurn, M.R. Andrews, A.P. Chikkatur,  S. Inouye, H.-J. Miesner, J. 
Stenger, and W. Ketterle}
\address{Department of Physics and Research Laboratory of Electronics, \\
Massachusetts Institute of Technology, Cambridge, MA 02139}
\date{Received November 5, 1997} \maketitle

\begin{abstract}
Bose-Einstein condensates of sodium atoms have been confined in an optical dipole trap
using a single focused infrared laser beam.  
This eliminates the restrictions of magnetic traps for further studies of
atom lasers and Bose-Einstein 
condensates.  More than five million condensed atoms were transferred into the optical trap.
Densities of up to $3 \times 10^{15} \text{cm}^{\text{-3}}$ of Bose condensed atoms were obtained,
allowing for a measurement of the three-body decay rate constant for sodium condensates
as $K_3 = (1.1 \pm 0.3) \times 10^{-30} \text{cm}^6 \text{s}^{-1}$.
At lower densities, the observed 1/e lifetime was more than 10 sec.
Simultaneous confinement of Bose-Einstein condensates in several hyperfine states was demonstrated.
\end{abstract}
]
\vskip1pc
The recent realization of Bose-Einstein condensation~\cite{ande:95,davi:95,brad:97} and of an atom laser~\cite{andr:97a,mewe:97} have sparked many theoretical and experimental studies of coherent atomic matter~\cite{bose:97}.
Yet, these studies are limited by the magnetic traps used by
all experiments so far.
For example, in the first demonstration of an atom laser, coherent atomic pulses were coupled
out into an inhomogeneous magnetic field, which served to confine the remaining
condensate.
Thus, during propagation, the pulses were exposed to Zeeman shifts.
While these shifts were mitigated by producing $m_F=0$ atoms, quadratic Zeeman
shifts may preclude precision experiments on such pulses.
Magnetic trapping also imposes limitations on the study of Bose-Einstein condensates,
since only the weak-field seeking atomic states are confined.  Since the atomic ground 
state is always strong-field seeking, weak-field seeking states can inelastically scatter into 
the ground state (dipolar relaxation) resulting in heating and trap loss.  Furthermore, trap 
loss is dramatically increased through spin relaxation collisions when different hyperfine 
states are simultaneously trapped, restricting the study of coherent or incoherent 
superpositions of different hyperfine states and the dynamics of multicomponent 
condensates.   Although in $^{87} \text{Rb}$ this increase is less dramatic due to a 
fortuitous cancellation of transition amplitudes~\cite{myat:97}, spin relaxation is still the dominant 
decay mechanism for double condensates.

All these problems are avoided if Bose-Einstein condensation is achieved in an optical trap 
based on the optical dipole force which confines atoms in all hyperfine states.  This has 
been one motivation for the development of sub-recoil cooling techniques~\cite{aspe:88,kase:92}, the 
development of various optical dipole traps~\cite{chu:85,phil:92,mill:93,take:96,kuga:97} and for pursuing Raman cooling~\cite{lee:96,kuhn:96}
and evaporative cooling\cite{adam:95} in such traps.
The highest phase space density achieved by purely optical means
was a factor of 400 below that required for Bose-Einstein condensation~\cite{lee:96}.
In this paper, we report the successful optical trapping of a Bose-Einstein condensate using 
a different approach:  first evaporatively cooling the atoms in a magnetic trap, 
and then transferring them into an optical trap.

This approach circumvents many difficulties usually encountered with optical dipole traps.  
Since the temperature of atoms is reduced through rf evaporation by a factor of 100, only 
milliwatts of laser power are needed as compared to several watts used to directly trap
laser-cooled atoms.  This ameliorates trap loss from heating processes in an optical dipole trap which are proportional to laser 
power, such as off-resonant Rayleigh scattering, and heating 
due to fluctuations in the intensity and position of the laser beam~\cite{sava:97}.  Furthermore, since 
the cloud shrinks while being cooled in the magnetic trap, the transfer efficiency into the 
small trapping volume of an optical dipole trap is increased.

The experimental setup for creating Bose-Einstein condensates was similar to our previous work~\cite{mewe:96a,andr:97b}.
Sodium atoms were optically cooled and trapped, and transferred into a 
magnetic trap where they were further cooled by rf-induced evaporation\cite{kett:96,walr:96}. 
The transition point was reached 
at densities of $\sim 1\times 10^{14} \text{cm}^{-3}$ and
temperatures of 1 -- 2 ~$\mu \text{K}$.
Further evaporation produced condensates containing 5 -- $10\times 10^6$ atoms in the 
$F=1,m_F=-1$ electronic ground state.  The atom clouds were 
cigar-shaped with the long axis horizontal, due to the anisotropic trapping potential of the 
cloverleaf magnetic trap, and had a typical aspect ratio of 15.

The optical trap was formed by focusing a near-infrared laser beam into the center of 
the magnetic trap along the axial direction.  For this, the output of a diode laser operating at 
985 nm was sent through a single-mode optical fiber and focused to a spot with a
beam-waist parameter $w_0$ ($1/e^2$ radius for the intensity) of about 6~$\mu 
\text{m}$.  This realized 
the simple single-beam arrangement for an optical dipole trap~\cite{chu:85,phil:92,mill:93,take:96}.
The infrared laser focus and the atom cloud were overlapped
in three dimensions by imaging both with a CCD camera.
It was necessary to compensate
for focal and lateral chromatic shifts of the imaging system which were measured using
an optical test pattern illuminated either at 589 or 985 nm.

The parameters of the optical trapping potential are characterized by the total laser power $P$ 
and the beam-waist parameter $w_0$.
The trap depth is proportional to $P/w_0^2$.
For a circular Gaussian beam, the trap depth is 1 
$\mu \text{K} / \text{mW}$ for $w_0 = 6\, \mu \text{m}$~\cite{depth_footnote}.
For such a beam, one expects an aspect ratio of the atom cloud of 27, with a geometric mean trapping
frequency $\bar{\nu}$ of
670 Hz at $P$ = 4 mW.
The measured frequencies of our optical trap were about half the expected values,
presumably due to beam quality.
Finally, due to the large detuning,
the spontaneous decay rate is small, leading to an estimated loss rate of one atom per 400 seconds.

Condensates were transferred into the optical trap by holding them in a steady magnetic trap while
ramping up the infrared laser power, and then suddenly switching off the magnetic trap.
A ramp-up time of 125 ms was chosen as slow enough to allow for adiabatic transfer, yet fast enough
to minimize trap loss during the ramp-up
due to high densities in the combined optical and magnetic traps.
Transfer efficiency was optimized for a laser power of about 4 mW,
with a measured mean trapping frequency $\bar{\nu} = 370$ Hz (see Eq. \ref{nu_equation}).
The transfer efficiency  dropped for higher laser power due to trap loss 
during the ramp-up, and decreased rapidly for smaller laser power due to the smaller
trap depth.
The sudden switch-off of the magnetic fields was necessitated by imperfections
in the trapping coils which displaced the center of the magnetic trap during a slow switch-off.
This limitation can be overcome in the future with auxiliary steering coils.

After a sudden switch-off of the optical trap, the freely expanding cloud was observed
after 40 msec time-of-flight using 
absorption imaging (Fig. 1).  The strong anisotropic expansion is characteristic of
Bose-Einstein condensates in strongly anisotropic trapping potentials.
Transfer efficiencies of up to 85\% were observed.

\begin{figure}
\epsfxsize = 7cm
\centerline{\epsfbox[156 282 395 448]{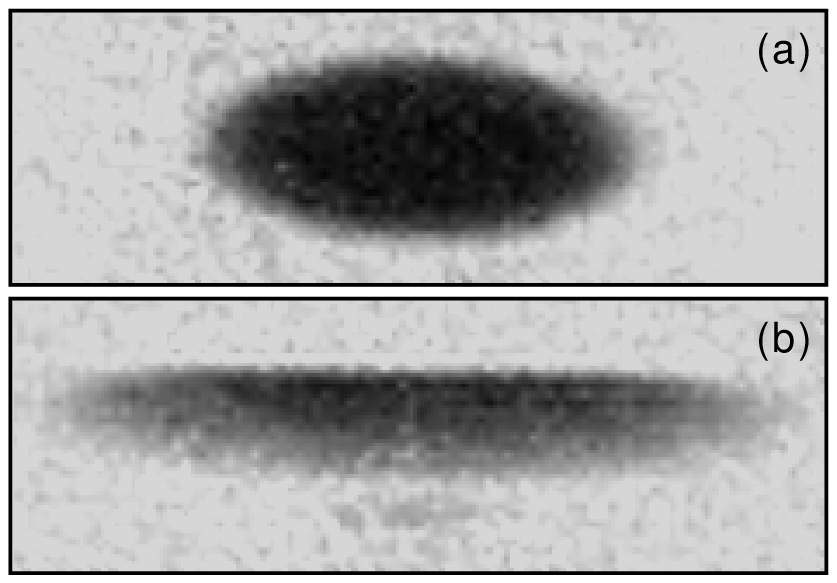}}
\begin{caption}
{Absorption images of expanding Bose-Einstein condensates, released (a) from the magnetic 
trap, and (b) from the optical trap with a mean trapping frequency of 370 Hz at $\sim$4 mW
infrared power.
The faster 
expansion in (b) is indicative of the higher densities of the optical trap.
The time-of-flight was 40 msec.  The field of view for each image is 2.2 by 0.8 mm.}
\end{caption}
\end{figure}

By loading the optical trap with magnetically cooled atoms at higher temperatures
and lower densities than those used in Fig. 1, we observed the sudden onset of a dense, 
low energy core of atoms amidst a broad background of non-condensed atoms (Fig. 2).
These data were obtained after 1 msec time-of-flight.  Hence,
the observed bimodality reflects the bimodal spatial distribution of atoms
in the optical trap, as opposed to the bimodal velocity distribution observed in previous
studies to verify the presence of a Bose-Einstein
condensate~\cite{ande:95,davi:95,mewe:96a}.
Two aspects are worth noting.
First, the number of thermal atoms is quite small due to the
small trapping volume and shallow trap depth of the optical trap which leads to a very small transfer
efficiency for thermal atoms.
The number of thermal atoms at the observed transition was measured at
24,000, which agrees quantitatively with a prediction based on the observed
trap depth and trapping frequencies, and the assumption that the thermal atoms
arrive at a temperature 1/10 of the trap-depth by evaporation.
This small upper limit contrasts sharply with the trajectory across the Bose-Einstein
condensation phase transition observed in magnetic traps,
where the number of non-condensed atoms at the transition temperature is much larger than the largest
number of condensate atoms eventually produced~\cite{mewe:96a,jin:97}.
Second, condensates were observed in the optical  trap in spite of its being loaded with
non-condensed magnetically trapped atoms.
This is due to the increase of phase space density
during the adiabatic process of ramping up the laser power~\cite{pins:97}.
A detailed study of this effect will be reported elsewhere.

\begin{figure}
\epsfxsize = 7cm
\centerline{\epsfbox[-2 10 253 179]{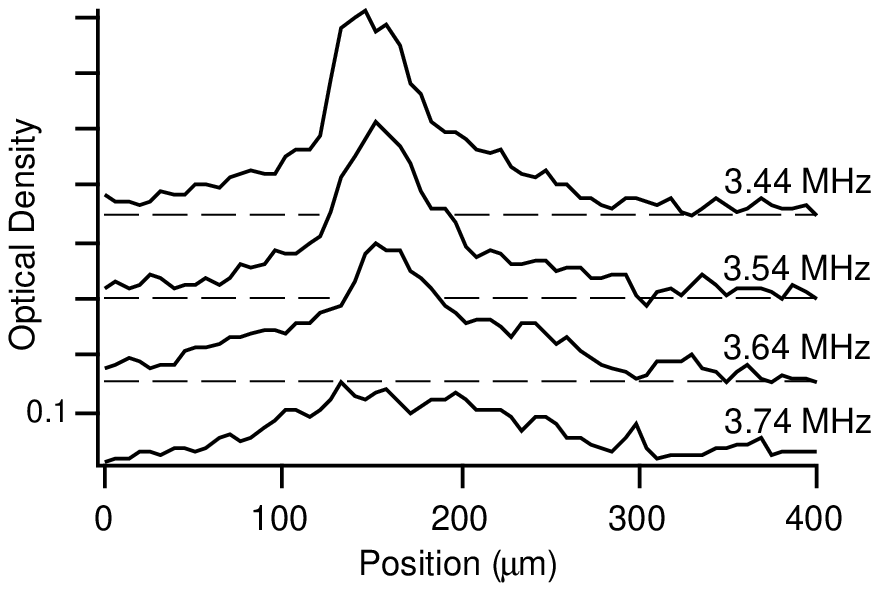}}
\begin{caption}
{Optical density profiles of optically trapped atoms.
Due to the short time-of-flight (1 msec), the profiles show the spatial distribution
along the long-axis of the dipole trap.  Profiles are labeled by
the final rf used in the evaporative cooling cycle.  The sudden bimodality observed
below 3.74 MHz indicates the onset of Bose-Einstein condensation.}
\end{caption}
\end{figure}

After the trap is switched off, the internal repulsive (mean-field) energy of 
the condensate is transformed into kinetic energy of the expanding cloud.  This allows for the 
determination of peak densities $n_0$ from time-of-flight data~\cite{mewe:96a}.  For a harmonic trapping potential 
in the Thomas-Fermi approximation, the spatial distribution is parabolic, both initially and 
during time-of-flight~\cite{cast:96}.  The average mean-field energy per atom is  $2/7 \, n_0 
\tilde{U}$, where $\tilde{U}=4\pi \hbar^2 a/m$ is proportional to the scattering length $a=2.75 \,\text{nm}$~\cite{ties:96}.
Assuming a predominantly radial expansion, the peak density was determined from
time-of-flight images by $n_0 \tilde{U} = m v_{max}^2 / 2$, where
$v_{max}$ is the ratio of the
maximum flight distance $\Delta l$ and the expansion time $\Delta t$.
The number of condensed atoms $N$ was measured by integrating the optical density in time-of-flight images.
The mean trapping frequencies $\bar{\nu}$  are related to $N$ and $n_0$ by~\cite{mewe:96a}
\begin{equation}
\bar{\nu} = 0.945\frac{\hbar \sqrt{a}}{m}n_0^{5/6} N^{-1/3}. \label{nu_equation}
\end{equation} 

The initial density of condensates in the optical trap was varied by transferring the atoms at settings
which maximized the initial transfer efficiency (see above), and then ramping the infrared power by a factor of two up or down in the all-optical trap. The infrared power was then kept constant for lifetime studies.
The peak densities achieved in this manner ranged from $3\times 10^{14} \text{cm}^{-3}$ in the
weakest optical trap
to $3\times 10^{15}\text{cm}^{-3}$ in the tightest.  For the lowest infrared power used, atoms were observed
spilling out of the optical trap, indicating that the depth of the trap was comparable to the 200 nK
mean-field energy of the condensate which remained.

The lifetime of condensates was studied by measuring the number of condensed atoms in
time-of-flight images after a variable storage time in the optical trap.
Results are shown in Fig. 3 for two settings of the infrared power, and also for the magnetic trap.  
The lifetime in the magnetic trap is very short unless 
the trap depth is lowered by ``rf shielding''~\cite{mewe:96a,burt:97}, allowing
collisionally heated atoms to escape.
Similarly, the long lifetimes observed in the optical trap are made possible by its
limited trap depth.
The observed loss rates per atom in the optical trap ranged from 4 $\text{s}^{-1}$  at a peak density $n_0=3\times 10^{15}\text{cm}^{-3}$ to less than 1/10 $\text{s}^{-1}$ at $n_0=3\times 10^{14}\text{cm}^{-3}$.

\begin{figure}
\epsfxsize = 7cm
\centerline{\epsfbox[6 2 241 179]{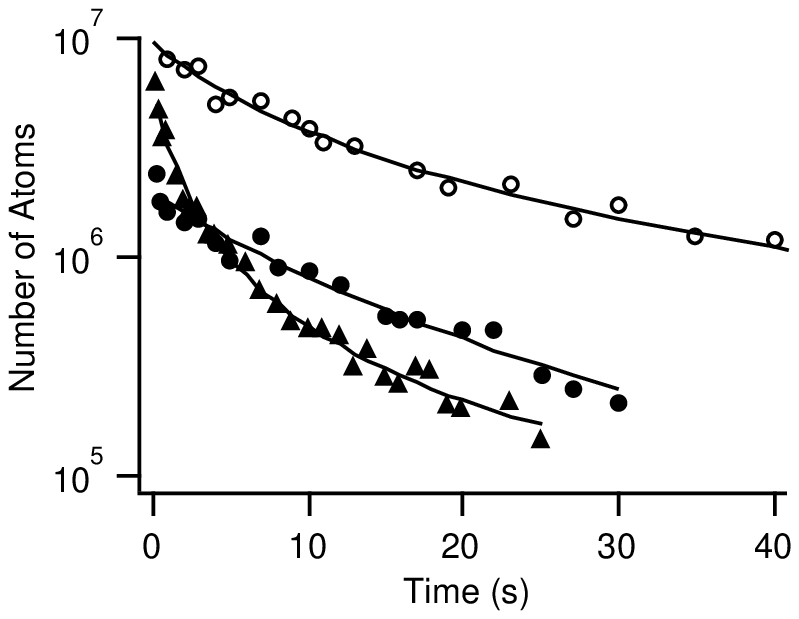}}
\begin{caption}
{Number of condensed atoms vs. trapping time.  Closed circles and triangles
represent data for the optical traps with the best transfer efficiency (370 Hz, $\sim$4 mW) and
the slowest decay (weakest trap, $\sim$2 mW), respectively.
Open circles represent data for the rf-shielded magnetic trap.
Error in the number measurements is estimated as 10\%.
Lines are
fits based on Eq. \ref{lossrate}.}
\end{caption}
\end{figure}

The decay curves in Fig. 3 are described by
\begin{equation}
\frac{dN}{dt} = -K_1 N - K_3 N <n^2>, \label{lossrate}
\end{equation}
where $K_1$ accounts for density independent
loss processes such as residual gas scattering, Rayleigh scattering and other external heating processes,
and $K_3$ is the rate constant for three-body decay.
The mean squared density $<n^2>$ can be derived from the peak density by
$<n^2>=8/21 \, n_0^2$.

Three-body decay was found to be the dominant loss mechanism in both the optical and the magnetic trap. By fitting the solution of Eq. \ref{lossrate} to the decay curves for the various optical
traps we obtained $K_1=(0.03\pm 0.02)$ $\text{s}^{-1}$
and
$K_3=(1.1\pm 0.3)\times 10^{-30} \text{cm}^6 \text{s}^{-1}$. This three-body decay
rate constant for $^{23}$Na is a factor of five smaller than for $^{87}$Rb~\cite{burt:97},
and can be ascribed completely to collisions among condensed atoms due to the
small number of non-condensed atoms in the optical trap.
Our result lies between two theoretical predictions of $K_3=3\times 10^{-29} \text{cm}^6 \text{s}^{-1}$~\cite{moer:96b} and $K_3=3.9\hbar a^4/2m=3\times 10^{-31}\text{cm}^6 \text{s}^{-1}$~\cite{fedi:96}. 
The loss rate due to dipolar relaxation (two-body decay) was predicted to be negligible at
the densities considered~\cite{moer:96a}.
While the decay curves show three-body decay to be the dominant loss mechanism,
they do not exclude two-body decay rates comparable to $K_1$.

One major advantage of the optical trap over magnetic traps is its ability to confine atoms in 
arbitrary hyperfine states.  To demonstrate this, the atoms were put into a superposition of 
F=1 hyperfine states by exposing them to an rf field which was swept from 0 to 2 MHz in 2 msec. 
Parameters were chosen in such a way that the sweep was neither adiabatic, nor diabatic,
similar to our work on the rf output coupler~\cite{mewe:97}.  The distribution over hyperfine 
states was analyzed through Stern-Gerlach separation by pulsing on a  magnetic field gradient of a few G/cm
during the 40 msec time-of-flight.  Fig. 4 demonstrates 
that all three states can be optically trapped.
By extending the time between the rf sweep and the probing, we confirmed that all F=1 hyperfine
states were
stored stably for several seconds.

\begin{figure}
\epsfxsize = 7cm
\centerline{\epsfbox[173 283 363 447]{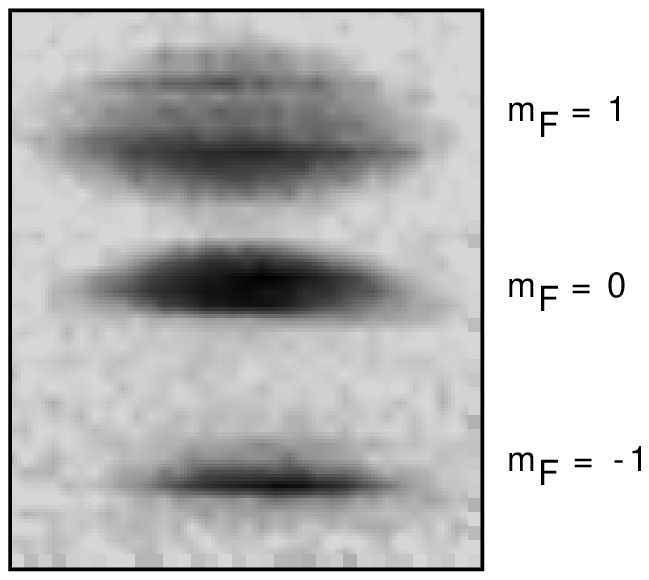}}
\begin{caption}
{Time-of-flight image of Bose-Einstein condensate of several hyperfine states.
An rf sweep was applied 90 msec before releasing the atoms from the optical
trap.  Hyperfine states were separated by a magnetic field gradient pulse
during the 40 msec time-of-flight.
All F=1 magnetic sublevels are visible.
The probe beam was $\sigma^{-}$ circularly polarized.
The field of view is 1.6 by 1.8 mm.}
\end{caption}
\end{figure}

In conclusion, we have realized an optical trap for Bose-Einstein condensates.  Due to the 
low energy of the condensates, just milliwatts of far-detuned laser radiation were sufficient to 
provide tight confinement.  More than five million condensed atoms were 
trapped, and lifetimes comparable to those in our DC magnetic
trap were observed.  Densities of $3\times 10^{15} \text{cm}^{-3}$ were
achieved, unprecedented for both Bose condensates and for optically-trapped atomic clouds.
High densities and high condensate fractions allowed for a determination of the
three-body decay rate constant in sodium as $K_3=(1.1\pm 0.3)\times 10^{-30} \text{cm}^6 \text{s}^{-1}$.
This trap offers many new opportunities to 
study Bose-Einstein condensates and atom lasers.  Since the optical trap works at arbitrary 
external magnetic fields, Feshbach resonances in the scattering length~\cite{ties:93} can now be 
observed for strong-field seeking states.  One can study coherence and 
decoherence of superpositions of magnetic hyperfine sublevels in the presence of
all spin-conserving collisions, and,
since the spin degree of 
freedom is no longer constrained by magnetic trapping, it may be possible to study spin waves
\cite{spin_articles} in a Bose-condensed gas.
The shallow and well controlled trap depth allows for new output-coupling schemes~\cite{raman}.
Finally, the optical trap may also serve as an ``optical tweezers'' to move condensates, and,
for example, place them in microcavities and on surfaces.

We are grateful to Dr. Gunther Steinmeyer, Eric Thorne and Prof. Erich Ippen of MIT for their help 
with the infrared laser.  This work was supported by the Office of Naval Research, NSF, 
Joint Services Electronics Program (ARO), and the David and Lucile Packard Foundation.  
A.P.C. would like to acknowledge support from an NSF Graduate Research Fellowship, J.S. 
from the Humboldt Foundation, and D.M.S.-K. from the JSEP Graduate Fellowship 
Program.


\begin{thebibliography}{10}
\bibitem{ande:95}
M.~H. Anderson {\it et~al.}, Science {\bf 269}, 198 (1995);

\bibitem{davi:95}
K.B. Davis {\it et~al.}, Phys. Rev. Lett. {\bf 75}, 3969 (1995);

\bibitem{brad:97}
C.C. Bradley, C.A. Sackett, and R.G. Hulet, Phys. Rev. Lett. {\bf 78}, 985 (1997), see
also: C.C.Bradley {\it et~al.}, Phys. Rev. Lett. {\bf 75}, 1687 (1995).

\bibitem{andr:97a}
M.R. Andrews {\it et~al.}, Science {\bf 275}, 637 (1997).

\bibitem{mewe:97}
M.-O. Mewes {\it et~al.}, Phys. Rev. Lett. {\bf 78}, 582 (1997).

\bibitem{bose:97}
Workshop on Bose-Einstein condensation, Castelvecchio, Italy, July 12-17 1997, 
Book of Abstracts.

\bibitem{myat:97}
C.J. Myatt {\it et~al.}, Phys. Rev. 
Lett. {\bf 78}, 586 (1997).

\bibitem{aspe:88}
A. Aspect {\it et~al.}, Phys. Rev. Lett. {\bf 61}, 826 (1988).

\bibitem{kase:92}
M. Kasevich and S. Chu, Phys. Rev. Lett. {\bf 69}, 1741 (1992).

\bibitem{chu:85}
S. Chu, J.E. Bjorkholm, A. Ashkin, and A. Cable, Phys. Rev. Lett. {\bf 57}, 314 
(1986).

\bibitem{phil:92}
W.D. Phillips, in {\it Laser Manipulation of Atoms and Ions}, edited by E. Arimondo, 
W.D. Phillips, and F. Strumia, Proceedings of the International School of Physics ``Enrico 
Fermi'', Course CXVIII (North-Holland, Amsterdam, 1992) p. 289.

\bibitem{mill:93}
J.D. Miller, R.A. Cline, and D.J. Heinzen, Phys. Rev. A {\bf 47}, R4567 (1993).

\bibitem{take:96}
T. Takekoshi and R.J. Knize, Optics Lett. {\bf 21}, 77 (1996).

\bibitem{kuga:97}
T. Kuga {\it et~al.}, Phys. Rev. Lett. {\bf 78}, 4713 (1997).

\bibitem{lee:96}
H.J. Lee {\it et~al.}, Phys. Rev. Lett. {\bf 76}, 2658 (1996).

\bibitem{kuhn:96}
A. Kuhn, H. Perrin, W. H\"{a}nsel, and C. Salomon, in {\it Ultracold Atoms and Bose-
Einstein-Condensation}, edited by K. Burnett, OSA Trends in Optics and Photonics 
Series, Vol. 7 (Optical Society of America, Washington D.C., 1996) p. 58.

\bibitem{adam:95}
C.S. Adams {\it et~al.}, Phys. Rev. Lett. 
{\bf 74}, 3577 (1995).

\bibitem{sava:97}
T.A. Savard, K.M. O'Hara, and J.E. Thomas, Phys. Rev. A {\bf 56}, R1095 (1997).

\bibitem{mewe:96a}
M.-O. Mewes {\it et~al.}, Phys. Rev. Lett. {\bf 77}, 416 (1996).

\bibitem{andr:97b}
M.R. Andrews {\it et~al.}, Phys. Rev. Lett. {\bf 79}, 553 (1997)

\bibitem{kett:96}
W. Ketterle and N.J. van Druten, in Advances in Atomic, Molecular, and Optical 
Physics, vol. 37, edited by B. Bederson and H. Walther (Academic Press, San Diego, 
1996) p. 181.

\bibitem{walr:96}
J.T.M. Walraven, in {\it Quantum Dynamics of Simple Systems}, edited by G.L. Oppo, 
S.M. Barnett, E. Riis, and M. Wilkinson (Institute of Physics Publ., London, 1996) p. 
315.

\bibitem{depth_footnote}
We assumed a two level atom.  25\% of the trap depth comes from the ``counter-rotating''
term usually neglected in the rotation-wave approximation.

\bibitem{jin:97}
D.~S. Jin {\it et~al.}, Phys. Rev. Lett. {\bf 78}, 764 (1997).

\bibitem{pins:97}
P.W.H. Pinske {\it et~al.}, Phys. Rev. Lett. {\bf 78}, 990 (1997).

\bibitem{cast:96}
Y. Castin and R. Dum, Phys. Rev. Lett. {\bf 77}, 5315 (1996).

\bibitem{ties:96}
E. Tiesinga {\it et~al.}, J. Res. Natl. Inst. Stand. Technol. {\bf 101}, 505 (1996).

\bibitem{burt:97}
E.A. Burt {\it et~al.}, Phys. Rev. Lett. {\bf 79}, 337 (1997).

\bibitem{moer:96b}
A.J. Moerdijk, H.M.J.M. Boesten, and B.J. Verhaar, Phys. Rev. A {\bf 53}, 916 
(1996).

\bibitem{fedi:96}
P.O. Fedichev, M.W. Reynolds, and G.V. Shlyapnikov, Phys. Rev. Lett. {\bf 77}, 
2921 (1996).

\bibitem{moer:96a}
H.M.J.M. Boesten, A.J. Moerdijk, and B.J. Verhaar, Phys. Rev. A {\bf 54}, R29 (1996).

\bibitem{ties:93}
E. Tiesinga, B.J. Verhaar, and H.T.C. Stoof, Phys. Rev. A {\bf 47}, 4114 (1993).

\bibitem{spin_articles}
B.~R. Johnson {\it et~al.}, Phys. Rev. Lett. {\bf 52}, 1508 (1984); P.~J. Nachter, G. Tastevin, M. Leduc, and F. Lal{\"o}e, J. Phys. (Paris) Lett. {\bf 45}, L411 (1984); W.~J. Gully and W.~J. Mullin, Phys. Rev. Lett. {\bf 52}, 1810 (1984).

\bibitem{raman}
G.M. Moy, J.J. Hope, and C.M. Savage, Phys. Rev. A {\bf 55}, 3631 (1997)

\end{thebibliography}
\end{document}